\documentstyle[12pt]{article}
\begin{document}
\hoffset = -1truecm
\voffset = -2truecm
\begin{centering}
{\huge\bf Where Has Entropy Gone: Theory of General System (II)}\\
\vskip .4cm
Zhen Wang\\
Physics Department, LiaoNing Normal University, Dalian 116029,  P.R.China\\
\vskip .7cm
{\bf Abstract}\\
\end{centering}
\vskip .4cm
A pair of symmetric expressions for the second law of thermodynamics  is  put
forward. The conservation and transfer of entropy is discussed and applied
to problems like biology, culture and life itself. A new explanation is given
to the cosmic expansion with the concept of diversity in this theory.
The problem of contingency and necessity
is also discussed.

\vskip .5cm
\noindent {\large\bf I.   Introduction}\\
\vskip .3cm
The world is a kaleidoscope. Both the lives in the world and the creations
of the lives are tremendous. For us human being, mankind is the greatest
life,
human culture is the most wonderful creation.
But if human culture could  not help us
transcend our limitation and enter the realm of freedom in a  broader
and higher sense, then it would not be great enough and
mankind would  be  no more intelligent than other animals. Today human being
has various  kinds  of culture and great amount of knowledge. But have we
obtained  an  altitude  at which we can have overlook at the various cultures,
and a  golden  string  to harmonically run through all knowledge ? In the first
paper I introduced  the theory of uncertainty quanta in a general system, as
well  as  some  of  its applications in some problems in physics and
mathematics. In  this  paper  we shall discuss another important part of
the theory of  general  system,   the conservation and transformation of order.
\vskip .5cm
\noindent{\large\bf II.     Discussion on the Second Law of Thermodynamics}\\
\vskip .3cm

The second law of thermodynamics is the most meaningful law in physics.
It is profound not only because it is a law that has got the most discussion
yet  a law that is most strange to us, but also because it gives us the
clue to  the understanding of life, i.e. the famous arrow of time.
For this famous arrow, physicists can at the most tell us that it is a
natural choice. Today we would not understand more about it without the
help of systematic view.

The second law points out that an isolated  system  will  evolve  in
such  a direction in which its entropy never decreases. This means
there is a special direction in our life. The time arrow points to the
direction  in  which  an isolated system gets more and more chaotic
and disordered until  its  entropy gets to its maximum. The entropy
is a variable indicating the disorder of the system. The bigger the
entropy is, the lower  the  level  of  order  of  the system is.
We call the order of a system its
negative entropy. In this theory,  the second law  embodies  the
limitation  of  the  observer  system.   After introducing the inversion
relation of system, we'll come back  again  to  the discussion of the
second law from a new angle (See the third  paper  ``Quantum Cosmology'' ).
Then the  unsymmetry of the second law will appear in a wonderful symmetry
relation. Here in this paper we shall only discuss it  through  the
relationship between different systems. So we express the second law in
such a way: a high-leveled (or highly ordered) system  may
deprive  the  negative entropy of low-leveled systems. In this theory,
the second law is only a possibility, not a necessity. It
is  the  reflection  of  the  relationship between different systems.
Thus in a sense, it is artificial.

Statistical physics tells us that the  increase  in  the
entropy  means  the increase in the microscopic weight of state of
the system. We  know  that there is a correspondence between the
states of a  system  and  that  of  its environment on any time
quantum. When we observe an isolated system evolving in accordance
with the second law of thermodynamics, the system turns
to  be the environment of the observer. Thus on any time
quantum of  the  observer,  there is a correspondence of the
state of the system with that of the observer. Actually, the
second law tell us such a thing: the observer (the system) will
correspond to such states of the isolated system (the environment)
that  get more and more microscopic weights. In other  word,
the  degeneracy  of  the observer system gets higher.
But what does this mean  ?   The  researches  of functional
material in recent decades has given us lots of inspiration.
Some special functional materials can select or discern the
polarity or  direction of free radicals at a distance.
The more ordered the functional material is,  the stronger
this selecting ability is. Just as we  discussed  in  the  first
paper ``System and Its Uncertainty Quanta'',  this  selecting  ability
is  the ability of a system to fix its environment. It decides
the abundance  of  its environment, which symbolizes
the degeneracy of the system. The degeneracy of a system and
the abundance of its environment  are  just  the  same  thing.
Therefore we see, a more ordered system can correspond to more
states of  its environment, thus has higher degeneracy.
So the second law of  thermodynamics actually reveals
that in the evolution of an isolated system,
the  observer gets more ordered while the isolated
system gets more disordered. Or you  can say, there is a
transfer of entropy.

Let's see how the second law works.
This is process which I call pararesonance of time quanta.
Imagine a system A, which is of high synergistic level,
and a system B, which is of relatively low level.
Then according to this theory, we have $t_{A}<t_{B}$.
As demonstrated in Figure 1, if $t_{B}$  is several times
bigger  than tA, then A will see naturally a number
of structures or possibilities,  which B can not discern,
in one time quantum of B. This makes it possible fr A
to affect the state of B at the will of A.
Like or  not,   B  has  to  face  the influence
that assimilate its time quantum to that
synchronous with  that  of the A system. There must
be the pararesonance because B is in the environment
of A (otherwise, the two system would be irrelevant
so that the two  systems do not exist relative to
each other. See the discussion about the four kinds
of relationship between systems in  ``Quantum Cosmology'' ) .
Such interaction happens in the basic structure of  space  and  time  and
it provides a basic background of space and time. So it does
exist. This is what I call the pararesonance. The time quantum
of B is likely to be synchronized with that of systems
that are of higher synergistic level. This means that it will
disintegrate and lose  its  independence  if  it  can  not  get
enough negative entropy from its own environment to sustain
its existence.  Then  it will have no independent time quantum
of its own, which is just the  case  of the second law of thermodynamics.

But here the most meaningful byproduct we get from the above discussion
is: If B can get enough negative entropy  to  sustain  its
existence,   then  such influence from higher leveled systems
is beneficial to its order  because  it tends to break up its time
quantum to smaller one. Moreover, when we think of the
extreme case in which A is the perfect system with zero time
quantum,  we  immediately  get  an  amazing conclusion that
such  tendency is in fact the basic property or structure  of finite
time quanta for all systems. Thus we may also express
the  second  law in the following form: a high-leveled (or  more  ordered)
system
may  give negative entropy to low-leveled systems. Of course it is still
nothing but a possibility. Now we have got two expressions for the second law
of
thermodynamics,  which seem to be in conflict. In  fact  it  is  this  conflict
that  embodies  the equality and symmetry of all systems in a higher sense.
The two expressions reveal two kinds  of  property  or nature
of a general system, which add to each
other to give us a  deeper  and more integrated
understanding of the world. I call the first the I expression
(Increasing Expression),  and  the  second  the  D  expression
(Decreasing Expression). Here many readers may come to the notion
that  the  D  expression will help us to understand
vast quantity of phenomena in chemistry, biology,  culture
and even society. We shall discuss the problem in the next
chapter.  Practically, absolutely isolated system does not
exist. The inner  environment of  a  system  is  not
closed  and  there  are  constant  interchanges  and
transformation between the  inner  and  outer
environment.   Therefore,   to summarize the two expressions
for the second law, we can get such  conclusion for a
general system: There may be entropy transfer between a system and
its environment.

Physicists always take the second
law  of  thermodynamics  as  an  infallible precept,
so that someone even declared
that a theory might still  be  correct if it violates other laws in physics,
but would be hopeless  if  it  did  not conform to the second law.
In order to find out the truth, we have to face the danger of being hopeless.
It  is  an  interesting  contrast  that  although scientists are so confident
about the second law,  they  can  not  provide  a harmonic and  unified  phys
ics  basis  for the  understanding  of  the  vast phenomena of
order in biology. Obviously, there must be a direction  of  time opposite
to the classical  second  law.   We  shall  find  out  the  opposite
direction of time in the third  paper "Quantum Cosmology". Here for the
time being,   we  shall  only  discuss  the limitation of
the second law from the
general property  of  and  relationship between a system and its environment.
During recent decades researches in non-equilibrium thermodynamics have ma
de  a  big  progress.   These  achievements support the expression I give
for the second law. The keypoint to  understand this is to renew our idea
for order. The nature of the order of a system,  or the negative entropy
as some physicists like to call it, is the degeneracy of
the system with respect to the states  of  its  environment.
More  ordered environment is farther away from equilibrium and
has  a  smaller  number  of microscopic states, thus the subject
system has a low degeneracy and low order.
 In such a case
the negative entropy has been transferred from the system  to its
environment. Once we have such an understanding of the second law, we
get the basis to renew our knowledge about this world.

Let's study the case in Fig.1  further. What would happen if
a lower-level system tries to imagine the environment of higher-level
system ?  Obviously a lower-level system can only discern part or even a
small part of the environment of the  higher-level system, which is
also a part of its own environment  and  seems  to  have  no unusual
significance to it. But the lower-level system will discover in its
environment that there are not only phenomena of super speed of
light but also violation of the second
law of thermodynamics (classical form).   Of  course
these may be out of the same reason.   There  is  only  one
one-to-one correspondence between a system and its environment on one time
quantum,  and different correspondences are unfolded on
different time quanta.   If
the  B system, which has faith for classical expression of the  second  law,
makes observation in its environment, it will  find  inexplicable  phenomena.
Why something or some state appear suddenly without intermediate course ?
Why the time can be reversed or transcend ? In Fig. 1 the
time quantum of B is  four times as large as
that of A, thus A can make four  choices  within  one
time quantum of B, of which B can't be aware. On the other hand,
when B is at  the time P, A may have seen or even
given some influence to the  time  Q  in  B's time scale. In such a case,
an incident of super speed of light happens to B.
So the future of B is seen or  influenced  by  A.
Because  such  influence happens in the uncertainty time quantum of
B, it can not  but  accept  it  as fate when the future becomes the present.
Reversely, when B is at the point Q,
A may have no difficulty in getting to the point P in B's history. Then
the second law of thermodynamics, the golden law of physics, is violated to B.

Physics has been generally regarded as the fundamental
subject underlying all other sciences.
But physicists have been long perplexed with the futility
in providing an integrated picture to understand the phenomena in biology,
let alone parapsychology and those unimaginable mysteries in nature.
Of  course the simplest and most efficient way to deal  with  those
inexplicable  in present theoretical frame is to denounce those as
pseudoscience deserving  no attention.
But if we want to get a higher altitude to understand  more,   we must
outstrip present theory, including the second law of thermodynamics. The
purpose of science is to relieve human being from ideological  barrier,   no
matter where the barrier comes from or whether it used to be beneficial to us.
\newpage
\noindent {\large\bf III.    The Conservation and Transfer of Entropy}\\
\vskip .3cm
The negative  entropy  symbolizes the order or the
synergistic level of a system.
When we talk about a system, we always have a corresponding
environment. Here the environment is referred to the
total environment including inner and outer environment.
There is  an one-to-one correspondence between the states of the
system and its environment. The more ordered the system,
the higher its degeneracy. This  means  it  can correspond
to more states of the environment in more efficient  ways.   There are many
such  examples  for  this  in  thermodynamics,   biology  and
other researches like the functional materials,
so it is not difficult to  come  to this point. But  here
there  is  no  reason  at  all  to  dissuade  us  from interchanging
the role of the system and the environment, i.e. to regard  the system
as the environment of its former environment. We  said  in  the
first paper "System and Its Uncertainty Quanta" that there
is some arbitrariness in delimiting a system and its environment, which
depends
on  the  synergistic function and our interest in the problem. Therefore the
system  and  its environment are born equal, and the correspondence between
them is  a  mutual one-to-one correspondence between two infinite sets.
The  system  can  affect the environment and vice versa. The system can  not
master  its  environment completely
because of its limitation, and the environment
has to be  affected by the system more or less.
A system can act on purpose or selectively,   but how can it deny with
reasonable logic that
the uncertainty in its environment comes out of some special purpose ?
If the system  could,   the  uncertainty would not be uncertainty any more.
In the third paper "Quantum Cosmology"
we shall have deeper understanding about the symmetric  relationship  between
a system and its environment. Here once we have enough spirit of  equality
and democracy to make the ideological breakthrough,  we  can  immediately
get  a profound relation: there is a complementary relation between
the
entropy  of the system and that of its environment. That is, the
higher the degeneracy of the system,
the lower the degeneracy of its environment. Or the more ordered the system,
the less ordered its environment. If the entropy of the system is designated
with S, and the entropy of its total environment S',   then  we
have

$$S + S^{\prime} = 0\eqno(1)$$

\noindent In fact, in the researches for the dissipative structure in recent
decades,  such entropy conservation has already dimly emerged. There, the
irreversible process that were formerly considered to generate disorder  has
become  good assistance for producing order,
order and disorder  are  less  hostile,   and disorder in some sense may
be preparation for order in a broader  sense.   We can see from (1) that if we
level  systems
according  to  their  synergistic functions, then order and
disordered together at any  level.   They are mutual and co-existent. In some
sense whether they are order or  disorder depends on how we see them. All
order  or  disorder  phenomena  make  up an inseparable and interwoven whole
together with ourselves. Evolution of things
is meaningless unless an environment is indicated. Environment is the content,
object and ways of being of the synergistic  function  of  a  system.   With
synergistic function, some order  is  transferred  from  the  system  to
its environment
 or reversely, which is in accordance with the expression  I  give for the
 second law of thermodynamics. Thus the order in the  environment  can be
 seen as to have come from its system. A system becomes the perfect  system
 once it achieves perfect harmony. Such perfect system has  infinite
 negative entropy and its environment has infinite entropy. So it is
 infinitely ordered and has an environment that is absolutely disordered,
 or you  may  say  that the perfect system has infinite selecting ability
 and an absolutely  obedient environment which, in fact, has completely
 merged into the  system.   But  in practical we can find no concrete
 system to be perfect system because  it  is much more superior to us
 present system. The perfect system has no environment. It has zero
 mass and time quanta but infinite space quantum. We see from (1) that
 when different  systems  become  the  perfect  system, they have no
 difference any more. They are totally the same. Yes, there is only
 one perfect system.

If we take a man as a system, we can have  a  better  understanding
in  this theory  about the difference among different people and
between  people  and other kinds of lives, or more generally,
other  systems.   Obviously, the difference is both inexorable
and infinite. When two systems  exist  in  the outer environments
of each other, then for any one of them,  the  other  does not
 exist. On the contrary, if two systems have enough common part
 in  their environments, they must be able to find some li
nkage between them. In such  a view, there must be some common
part among people, among all lives, and  even among all systems
that are known to us. But there are also endless differences. The
nature of the differences is that systems  with  different  orders
have different selecting abilities therefore different environments.
In fact it is not difficult to come to this conclusion. What is really
surprising  is  why it is so difficult for people to get rid of an
unreasonable belief  that  all people, even 
all lives live in a common, independent and complete environment
as is "commonly" sensed. Apparently this is  an  epistemological
limitation. Equality at low level and in small range conceals
inequality  at  high  level and in large range.

As an example of the theory of general system, let's
study the human  culture, which is something common and very
important for human being. The word culture here is referred to
the whole body of all kinds of special cultures. It is the linkage
among individuals, arteries and veins of the society. No matter what
form it may take, it is in its nature a relationship of life  and
surpasses all languages. Obviously culture is an order  phenomenon.
As  we  discussed above the order in culture can be seen as to have
come  from  the  order  of human being. Thus in its nature culture
is a phenomenon  of  life  that  has obtained negative entropy from
mankind.   In  this  sense,   culture  is  no different from other
animals on the earth. They are all assistants for  human being to
extract negative entropy from a wide background of  disorder.
They serve as storage of negative entropy for human being. This
is why the culture system, or any of its subsystems
like politics,  economy,   science,   art,  language a
nd etc., show some  characters  of  life  when  it  advances
  to  a relatively high level. The evolution of human society
  also shows features  of life. This has  resulted  in  the
  similarities  of  methodologies, modes  of development and
  basic difficulties in different areas. It is the same  reason
  that gives vitality to many frontier and cross disciplines,
  just like a  life develops a new organ or advances a new kind
  of function in a new environment.

But on another hand the culture system is significantly different
from animal systems because it does not have mass, time and space
quanta. Instead,   its uncertainty quanta describe other properties
than mass, time and  space,   or they are in different spaces in
popular jargon. That's why it does not have a visible and independent
physical body like plants  and  animals,   but  only exists in people's
participation. So it is a completely parasitic  life.  It reveals the
origin of its negative entropy
more  clearly  than  other  order phenomena. All the plants and
animals  draw  order  from  the  vast  disorder background in the
most efficient ways for themselves and  thus  for  us  human being
(remember, order means degeneracy in states).  They  have  really
 been part of the life of human  being.   In  this  respect  culture
 is  far  less efficient and helpful. It has both helpful and harmful
 effects. In some cases, it is just the harmful  ingredient  of
 culture  that  makes  mankind  feel "hungry"
and
then appease its hunger with its own body. In this theory, because
of the one-to-one correspondence between a system and its environment,
   the deterioration of our environment well embodies the withering of
   life  of  the mankind. Compared with real life in our environment,
   culture is  a  parasitic and low calibre life in its efficiency
   and harmonization for human being.

The function of culture has always been a controversial topic. Of
course there would not be my present paper if there were no culture.
For the gigantic system of the present human society, no social
progress can be  made  without the help of culture. But order
does not have only one form  to  take.   ( 1)  reminds  us that
culture is neither the aim nor a mark of human progress.  It is
only a tool in our way to perfection.
\vskip .5cm
\noindent {\large\bf IV.     Diversity and Uncertainty Quanta}\\
\vskip .3cm
In this theory, diversity of the environment is also an important
concept symbolizing the level of order for a system. It is equivalent
to an uncertainty quantum. They are the two sides of one coin. Diversity
  is  the  abundance  of  existence  in  the environment of a system, or
  equivalently the selecting ability of the system. Apparently, diversity
  is the relative variable of the  environment  for  the degeneracy of the
  system. The higher  degeneracy  a  system  has,   the  more abundant its
  environment is, or the richer its diversity. From the  point  of
  uncertainty quanta we can also get a view on diversity. Smaller
  mass  quantum means smaller basic brick for our physical reality,
  and  thus  more  abundant forms of existence. We know that mass is
  closely related to energy, and  mass quantum is a mark of the energy
  scale for system.  Therefore  we  can  say that system with smaller
  mass quantum is of higher level of energy,   or  has stronger selecting
   ability, thus it can have more  choices
  or  get  to
more details in fixing its environment. So companied with the diversity
is  always a due selecting ability. Here we see again  the  same  reasoning
  as  in  the mathematical consideration (See the first paper "System and
  Its  Uncertainty Quanta" ): infinite and infinitesimal, or rather, the
  up and low limits,  are related in a profound way. Therefore we see
  that  a  rich  diversity  of  the environment reveals a small basic
   unit, i. e.   uncertainty  quantum,   which indicates a strong
   synergistic function
and small uncertainty.

In such theory we can understand more deeply the implication of
extinction of the species on the earth. According to some experts,
a quarter of the present species will face danger of extinction in
about thirty years. We  know  that  lives  in  our  environment  are
 the  richest collection as well as the most efficient storage of
 forms of physical entity. Thus for human being, such an indefinable
 loss will never limit  its  harmful influence only within our present
 industry and agriculture, but immerse us
in a vast and threateningly clearing shroud of jeopardy. It has already
changed our future in a way of which we are still unaware. It's true that
we have advanced science and technology today, but we have lost biological
diversity in our world, which is far more valuable. This means  that  the
 synergistic level of human individuals, thus the whole mankind, has been
 lowed down,   or our human system is evolving in the way to disorder. You
 may  also  say  that the descent in our synergistic level has
resulted in the loss  of  biological diversity in our environment.
In recent years more and  more  public  concern has given to this
problem. The loss of a life in our environment means a loss of order
in our own life. What do we lose for the extinction of  species  in
our environment ?

The biological diversity is an appropriate indicator  for  human
synergistic level from systematic angle. From (1) we see that a more
 ordered system has a more disordered environment. Thus richer biological
 diversity  symbolizes  a more ordered system and a more disordered
 environment. But  here  is  a  very important but often misleading
  concept that needs explanation.  That  is  the concept of order.
  Why should a richer biological diversity designate  a  less ordered
  environment ? As a matter of fact,
life is  itself  a  phenomenon  of order in common sense. But to  the
observer  system,   it  may  be  seen  as concentrated manifestation
of disorder of the environment, for an environment with more abundant
biological diversity has more forms of existence, or more states. The
concept of order involved in (1) is based on a more  fundamental and
more general meaning.  It  is  the  abundance  of  states  of  the
basic particles (or rather, basic units) in the environment. The
richer  the  life phenomena
in
the environment, the more states the environment has,  thus  the
system has a higher degeneracy (corresponds to more environmental
states) and is more ordered. It is the order manifested in individual
 lives that creates the diversity of states of the whole physical
 reality,  or  disorder  of  the environment, which in turn corresponds
 to the order of the system. Here again we see the transformation and
  interweaving of  order  and  disorder.  This supports our new
  expression for the second law of thermodynamics and adds  to our
  understanding of the relation of  (1).

Such profound relation between uncertainty quantum  and  diversity
has  been embodied not only  in  mass  quantum,   but  also  in  time
quantum  in  the pararesonance, and in the linkage of empty set and
infinite in  mathematics.  Here we shall consider the counterpart in
space quantum, from  which  we  get very naturally the relation of
space quantum and the  cosmic  expansion.   Of course it is quite
common to have different explanations for a same phenomenon, which
all might be correct to some extent. I don't want
to deny the  success of other theories in cosmic problems.
  But in my opinion,  irregularity  still means the defect of the present
   theory unless  we  can  endow  some  physical meaning to it that is
   reasonably as well as logically acceptable.  In  this paper I just
   present a picture for the cosmic expansion. There are still some
   theoretical details that need to be worked out for other phenomena
   observed.

We have seen from the above discussion that some high calibre order
has  been turned into low calibre order in human system because of
the loss of negative entropy. Therefore the synergistic level of
human system in general has been lowed down (See also the third
paper "Quantum Cosmology" ). That  means  the mass and time quanta
are being enlarged while  the  space  quantum  is  being shortened.
We know that the space quantum is the smallest distance  in  which
all spacial points are equal. It is the basic 
unit of our space and  we  know nothing inside
the basic brick. So we have not the  least  reason
to  assume that inside the uncertainty quantum there
is nothing worth consideration  but a trivial series
convergent to limit zero. We should not make any  assumption
for the property of uncertainty quantum related to its inner
structure,   no matter how reasonable it might seem, because
that is beyond our comprehension according to the definition
of uncertainty quanta. As  a  matter  of  fact,  fractal
geometry has given very good examples of divergence. In
my  theory,  the expansion of the universe is just the
direct result of the  reduction  of the space quantum in human
system. The reduction of the space quantum implies the  reduction
 of  the  "length"  that  is  composed  of   equivalent   and
 indistinguishable points. As a result, some "distance" that
 was formerly  composed of  equivalent  and  indistinguishable
 points  becomes  unequivalent  and distinguishable. Thus there
 must  be  an  increase  in
the  visible  spacial distance. In other word, some visible
distance has been "produced"  from  the space quantum, the
basic unit of space, resulting in  the  expansion  of  the
space.

The cosmic expansion is the experimental  cornerstone  of
present  cosmology. What kind of cosmology will our brand-new
explanation for the phenomenon  lead to ? We shall have further
discussion for this problem  in  the  third  paper "Quantum
Cosmology".
\vskip .5cm
\noindent {\large\bf V.      Evolution of Life}\\
\vskip .3cm
There is a tremendous saying in philosophy that space and time are
the way of being for matter, which reveals the dependence of space
and time on mass. In fact it is only a part of a more profound relation.
A  specific  environment, which is described with three specific
uncertainty quanta, always belongs to a specific system. More ordered
system has more powerful  synergistic  function and thus corresponds
to more states of its environment, therefore  it  has  a more abundant
environment. There is an
one-to-one correspondence between  the states of a system and its
environment. In this sense, a man has no essential difference from
other forms of life. They all have some degeneracy and independence
relative to the environment. But what is the nature of life?

Scientific developments seem to have accumulated more and more
evidence  that all life activities have physical basis and are
ensured with  matter  in specific ordered form, and all spiritual
processes correspond to  some physical changes. But  i am afraid
that most people do not agree on such view of extreme reductionism.
People even have developed special science and  therapeutics for
spiritual behaviour. But such subject has never found in physics
its basis that can be widely accepted. People can say
nothing about the nature  of  spiritual  activity but that it
is a kind of function of the brain. Spiritual activity is one
of the most important features of life phenomena,  so we still
have a long way to go for the understanding of life. But here
is a prerequisite, i.e., we must first admit that the nature
of  life  phenomena can be understood. If there were something
in our world that we  would  never be able to understand, then
all the knowledge that human being  has  acquired would be of
no sense: a law would not be a law if it might fail at  any
time and nothing would exist in a world with
no laws.

In my theory, the diversity  and  uncertainty  are not  only
the  basic characteristic of life but also constitute the
essential part of it.  Systems with high synergistic  level
have  rich  environments.   A  system  and  its environment
have the relation revealed in (1).  Spiritual  activity  has
two meanings. One is the richness of diversity in environment.
Obviously, systems with  rich  diversities  have  high  degeneracy
according  to  our  above discussion, therefore are more ordered
and have rich
 spiritual activities. In this sense different life systems may
 have enormous differences in their spiritual behaviour because
 of the differences in their environments. On the other hand,
 it is uncertainty that more notably reminds  us the existence
 of spirit in daily activity. In fact it is  the  more  profound
 side of spirit. Uncertainty comes from the outer environment  of
 our  system and it reveals the limitation of our selecting ability.
 What is more important, in manifestation for our limitation it
shows  us  the  infinite  potential  of cognition we have, i.e., all
limitations can be realized by us in their nature. This potential is
the same for all life systems, which epitomizes the conservative
relation implied in (1). The correspondence relationship between
system  and its environment is the same in nature for all systems,
no matter high or low. In such a sense, spiritual and life  phenomena
all  acquire  some  kind  of absolute meaning. We shall have further
discussion about the relative and
the absolute meaning of life in the third paper "Quantum Cosmology" .

History tells us that although human society  has  always  developed
in  the direction toward wider and wider equality (it is so because
the society  also has life feature, according to my theory), it has
always been very  difficult for human being to get rid of a sense of
superiority. So whenever  scientific progress abolishes a special
advantageous status of mankind, it always  gives people a great
shock for a period. As is often seen,   this blind  sense  of
superiority usually accompanies the lack of
self-confidence. The development of computer gives a good example.
Today computers are so advanced  that  some people begin to worry
about the possibility that a race of computer may emerge and
threaten human being some day. In fact this is  impossible.
In  the  abundant diversity of human environment, there  is
enormous  amount  of  incalculable ingredient as well as
calculable ingredient. We  have  three uncertainty quanta
for mass, space and time, which gives us a very good sense
of consciousness
of our environment. Such sense of  consciousness  would  be
greatly different for an inorganic system, whose environment
is  too  simple because of its poor synergistic function.  No
matter  how  advanced  future computer technology will be, the
computer, in common sense, is only a  simple system with only
one uncertainty quantum for mathematics, mere  extension  of
human organ. Its physical structure is too simple to hold enough
negative  entropy for it to develop its own mass, space and  time
quanta
which
  are  delicate enough to make it alive. As a system with three
  uncertainty quanta in general sense, it is even less ordered
  than an ant. It is more preposterous to think that a race of
  computer would threaten the whole human being. What is more,
  can the  evolution  of  life  take  such  direct  route  that
  surmounts  the significant difference between organism and
  inorganism ?

Darwin's theory of evolution tells us  that  a  species
gradually  completes itself on its way of evolution through
natural selection. Environment plays a crucial role in this
 process. Variations are maintained  and  developed  when
 they suit the environment, diminished  and  eliminated
 when  doesn't.   This theory has achieved great success.
 But the mutation in this theory is  random incident that
 lacks explanation. In our theory, because of  the one-to-one
 correspondence  between  system and  its environment,  system
 fixes  its environment with  its  synergistic function, reversely,
 changes  in  the environment require appropriate variation of the
 system, no matter whether we can perceive the changes. Therefore
 all variations of the system are induced by some specific
 environment, and  they  embody  some  requirements  of  the
 environment. Without suitable environment a system can  not
 emerge and exist.   We  are constantly changing, so is our
 environment,
which in turn constantly  induces new systems to 
 realize the changes in a most efficient  way.
 Life  develops like this. Strictly speaking, there
 is no such thing as mutation. Even  birth and death,
 the special  mutations  of  life,   are  also  embodiment
 of  the requirement of changes in the relationship between
 system and its environment (See the third paper  "Quantum
 Cosmology"  ) .   The  word  mutation  still symbolizes the
 limitation in our knowledge. But if so, what causes change?

As a matter of fact, once we get further understanding of the
nature of  time, then evolution itself is also relative. The
development  in  some  function means increasing of order in
this respect. But doesn't the order increase  in one aspect
at the price of decreasing  in  another  aspect? Doesn't  it
increase in a small scope at the price of decreasing in a large
scope ?    We see from (1) that entropy can not be created but
only  flow  from  system  to environment or reversely, which
correspond to the 
expression I give  for  the second law of thermodynamics.
Therefore evolution of life should be appraised from its
total synergistic function, from its whole environment and
from all the relationship between it and  its  environment,
rather  than  partially concentrated on some particular
functions.

In recent decades studies in dissipative structure  have
given  us some  inspiration for the problem of origin of
life. Our environment is in constant change, so it may
become quite common for a system, away or  even  far
away from equilibrium, to appear. Thus it is imaginable
that  some  phenomena  of self organization may emerge under
some special conditions. But  the  problem of genesis of life
has not been completely solved.  Why  is  our  environment always
in change ? No satisfactory solution for  
the  problem  can  be  found without a profound
understanding of the nature of time. On what basis is
the symmetry of the I Expression and D Expression of the
second law established ? How and why did the universe
originate ? We should  have  an  integrated  and harmonic
theory to understand these questions, which we  shall  have
further discussion in the third paper "Quantum Cosmology" .
\vskip .5cm
\noindent {\large\bf VI.     Contingency and Necessity}\\
\vskip .3cm
Is this a world of contingency or necessity ? Is  everything
in  this  world ruled by probability or by a supreme adjudicator ?
Such questions have been a topic of dispute for philosophers for a
long time. According to this theory, different systems have different
environments. Thus any contingency and  necessity must be related  to
 a specific system and its environment, therefore relative. The
 contingency  and necessity in an environment reveal the synergistic
 level  of  the  system  to which the environment belongs.

Different systems may be of different  synergistic
levels  and  have  different,
even opposite aims for the selection of environment.
In such case, the environment will be fixed at the will of the
high-leveled system at the  price  of  some extra order,
because both systems
have to face the  increased  unsymmetry  in
their environments. As I said above,
this  fixation  is  relative  and
still affected with uncertainty revealing
the limitation.  Suppose  there  are  two
systems P and Q, with Q being of higher
synergistic level, and an incident
Y in  the  common  part  of  their  environment.
Thus  Y is related with both P and Q, though strictly speaking,
it may have different forms of existence in the two environments. If  their
aims  in  fixing  the environment are close or
in accordance with each other,  both  of  them  will save
order (remember, order
represents degeneracy). But  if  their  aims  are contradictory, Y will be
decided by Q, and both of them will lose some  order to balance the increased
order in their environments. Then the environment is necessitarian and
deterministic for Q but contingent and undecidable for P.  So we see that
it depends on the selecting ability or the  synergistic  level
of the system what roles contingency and necessity play in its  environment.

Strictly speaking, every imperfect system P may be in the  environment  of
a more highly-leveled system Q, which has smaller  mass  and  time  quanta
and larger space quantum. Thus Q can fix the environment of P system in P's
outer environment in a way that is imperceivable to P.
That is, when a high-leveled system observes a lower-leveled system in its
environment, it  will  clearly see the contingency in the lower-leveled
system, because for one state of the latter, the high-leveled system may have
several equivalent states to choose from. When the high-leveled system is the
perfect system with zero mass  and time quanta and infinite space quantum,
then  fortuity in  an  ordinary system is absolute and inevitable.
That is the fortuity in  an  imperfect system, because all imperfect systems
are in correlation with its environment in the level of the perfect system.

We may also have an understanding on this problem from another angle. As we
know, the state of a system can be described with three uncertainty quanta
which, as the name suggests,  also  reveal  the uncertainty in the
environment of the system. A system can not  perceive  the changes in its
outer environment, i.e., changes smaller than its mass and time quanta but
larger than its space quantum,  therefore  it  has  to  face  the results of
these changes without  knowing  the  reason.   So  there  must
be contingency in its environment. Limited space quantum restricts the range
of the synergistic function of the system. On any time quantum,   a  system
can only act on matter within  its  space  quantum,   which,   according
to  our discussion on correlation in the first  paper  "System  and  Its
Uncertainty Quanta", is actually correlated to all the matter in the
entire environment.

It is the same with human world. Contingency and necessity coexist. This means
 there are still things that people can not fix or control.  But  there  have
 never been earnest logic or conclusive proof to show that people can not
 get rid of fortuity in their environments. Where has the contingency in
 human world originated ? Obedience  to  fate  as  well  as  blind
 arrogance  often seriously restrict our thinking on this question. In
 this theory, a system with limited uncertainty quanta is doomed to have
fortuity in its environment. Though fortuity can not be avoided in daily
life for ordinary people, it may be quite different during different
periods or among different people. In fact, human history is the records
of victories over  contingency,  in which the developments of science are
the milestones of human emancipation. As the best embodiment of the human
spirit in pursuit for  truth,   science contains the most positive factors
of life in a profound way.   Einstein  was correct. God never plays dice. 
But unfortunately we are not God. Science today has not given us His
omnipotence yet. Then can a system really  get  to  the perfect state in
which the system has zero mass and time quanta but  infinite space quantum,
so that thoroughly wipes out contingency ? The answer of  this theory is yes.
That is the very important concept of  perfect  system  in  our theory, which
is in fact the starting point for us to understand the  world.  We shall
discuss it in the third paper "Quantum Cosmology" .
\end{document}